\documentclass[12pt,preprint]{elsarticle}

\usepackage{graphicx}
\usepackage{amssymb}
 \newcommand{\be}[1]{\begin{equation}\label{#1}}
 \newcommand{\ee}{\end{equation}}

 \journal{Physics Letter B}
 \begin{document}
 \begin{frontmatter}
\title{Exact solutions for shells collapsing towards a pre-existing black hole}

\author[L1]{Yuan Liu}\ead{yuan-liu@mails.tsinghua.edu.cn}
\author[L1,L2,L3]{Shuang Nan Zhang}\ead{zhangsn@mail.tsinghua.edu.cn} \ead{zhangsn@uah.edu}
\address[L1]{Physics Department and Center for Astrophysics, Tsinghua
University, Beijing, 100084, China}

\address[L2]{Key Laboratory of Particle Astrophysics, Institute of High
Energy Physics, Chinese Academy of Sciences, Beijing, China}
\address[L3]{Physics Department, University of Alabama in Huntsville, Huntsville, AL35899, USA}

\begin{abstract}
The gravitational collapse of a star is an important issue both for general relativity and astrophysics, which
is related to the well known ``frozen star" paradox. This paradox has been discussed intensively and seems to
have been solved in the comoving-like coordinates. However, to a real astrophysical observer within a finite
time, this problem should be discussed in the point of view of the distant rest-observer, which is the main
purpose of this paper. Following the seminal work of Oppenheimer and Snyder (1939), we present the exact
solution for one or two dust shells collapsing towards a pre-existing black hole. We find that the metric of the
inner region of the shell is time-dependent and the clock inside the shell becomes slower as the shell collapses
towards the pre-existing black hole. This means the inner region of the shell is influenced by the property of
the shell, which is contrary to the result in Newtonian theory. It does not contradict the Birkhoff's theorem,
since in our case we cannot arbitrarily select the clock inside the shell in order to ensure the continuity of
the metric. This result in principle may be tested experimentally if a beam of light travels across the shell,
which will take a longer time than without the shell. It can be considered as the generalized Shapiro effect,
because this effect is due to the mass outside, but not inside as the case of the standard Shapiro effect. We
also found that in real astrophysical settings matter can indeed cross a black hole's horizon according to the
clock of an external observer and will not accumulate around the event horizon of a black hole, i.e., no
``frozen star" is formed for an external observer as matter falls towards a black hole. Therefore, we predict
that only gravitational wave radiation can be produced in the final stage of the merging process of two
coalescing black holes. Our results also indicate that for the clock of an external observer, matter, after
crossing the event horizon, will never arrive at the ``singularity" (i.e. the exact center of the black hole),
i.e., for all black holes with finite lifetimes their masses are distributed within their event horizons, rather
than concentrated at their centers. We also present a worked-out example of the Hawking's area theorem.
\end{abstract}

\end{frontmatter}

\section{Introduction}\label{sec1}

The ``frozen star" is a well known novel phenomenon predicted by
general relativity, i.e. a distant observer ($O$) sees a test
particle falling towards a black hole moving slower and slower,
becoming darker and darker, and is eventually frozen near the event
horizon of the black hole. This process was vividly described and
presented in many popular science writings \cite{1,2,3} and
textbooks \cite{4,5,6,7,8,9}. The time measured by an in-falling
(with the test particle) observer ($O'$) is finite when the test
particle reaches the event horizon. However, for $O$ it takes
infinite time for the test particle to reach exactly the location of
the event horizon. In other words, $O$ will never see the test
particle falling into the event horizon of the black hole. It is
therefore legitimate to ask the question whether in real
astrophysical settings black holes can ever been formed and grow
with time, since all real astrophysical black holes can only be
formed and grown by matter collapsing into a ``singularity".

Two possible answers have been proposed so far. The first one is that since $O'$ indeed has observed the test
particle falling through the event horizon, then in reality (for $O'$) matter indeed has fallen into the black
hole. It seems that this paradox could be solved in the point of view of $O'$ or other comoving-like
coordinates. However, since $O$ has no way to communicate with $O'$ once $O'$ crosses the event horizon, $O$ has
no way to `know' if the test particle has fallen into the black hole. More importantly, in a real astrophysical
observation within a finite time, it is more meaningful to discuss this problem in the coordinate of $O$, which
is the main purpose of this paper. The second answer is to invoke quantum effects. It has been argued that
quantum effects may eventually bring the matter into the black hole, as seen by $O$ \cite{10}. However, as
pointed out by \cite{11}, even in that case the black hole will still take an infinite time to form and the
pre-Hawking radiation will be generated by the accumulated matter just outside the event horizon. (However, as
we shall show in this work, the in-falling matter will not accumulate around the event horizon.) It has also
been realized that, even if matter did accumulate just outside the event horizon, the time scale involved for
the pre-Hawking radiation to consume all accumulated matter (and thus make the black hole ``black" again) is far
beyond the Hubble time \cite{12}. Thus this does not answer the question in the real world. Apparently $O$
cannot be satisfied with either answer.

In desperation, $O$ may take the attitude of ``who cares?" When the
test particle is sufficiently close to the event horizon, the
redshift is so large that practically no signals from the test
particle can be seen by $O$ and apparently the test particle has no
way of turning back, therefore the ``frozen star" does appear
``black" and is an infinitely deep ``hole". For practical purposes
$O$ may still call it a ``black hole", whose total mass is also
increased by the in-falling matter. Apparently this is the view
taken by most people in the astrophysical community and general
public, as demonstrated in many well known textbooks
\cite{4,5,6,7,8,9,13} and popular science writings \cite{1,2,3}.
However when two such ``frozen stars" merge together, strong
electromagnetic radiations will be released, in sharp contrast to
the merging of two genuine black holes (i.e., Schwarzschild
singularities); the latter can only produce gravitational wave
radiation \cite{12}. Thus this also does not answer the question in
the real world. As we shall show in this work this view should be
changed, because in real astrophysical settings $O$ (in the
Schwarzschild coordinates) does observe matter falling into the
black hole.

This problem is closely linked to the problem of collapse. Great
interests have been paid on the problem of gravitational collapse.
Since the seminal work of Oppenheimer and Snyder (1939) \cite{14},
numerous works have been done on this issue. For example, the result
of \cite{14} has been extended to more realistic and complicated
cases, e.g. considering pressure, inhomogeneous, the effect of
radiation and rotation, and so on (e.g. \cite{15,16,17,18,19}).
There are also a lot of works concerning collapsing shells. The
paper of Isarel \cite{20} found the equation of motion for one thin
(without thickness) shell in the vacuum. Khakshournia and Mansouri
\cite{21} obtained the equation of motion for one shell in vacuum
with thickness up to the first order. Gon\c{c}alves \cite{22}
discussed the solution for two thin shells in vacuum. However, in
many real astrophysical settings, the matter shell collapsing
towards a black hole should have a finite thickness and is much less
massive than the black hole. In this paper we investigate such more
realistic cases, and in particular to find the differences between
these solutions and that in \cite{14}, and the implication for the
``frozen star" paradox. In Sec. \ref{sec2}, we present the dynamical
solutions for one and two shells, in which the shell is spherically
symmetrical and pressureless. We also show the evolution of event
horizon (the boundary of the region from where the photon cannot
escape to space-like infinity) and apparent horizon (the boundary of
the trapped region, see the detail of the definition of event and
apparent horizon in \cite{13} and \cite{23}) in the one shell case.
 In Sec. \ref{sec3}, we discuss the implications of these solutions and make
our conclusions. We adopt $G=c=1$ throughout this paper.

\section{The dynamical solution}\label{sec2}
\subsection{The solution for one shell}\label{sec21}
\begin{figure}
\begin{center}
\includegraphics[width=0.5\textwidth]{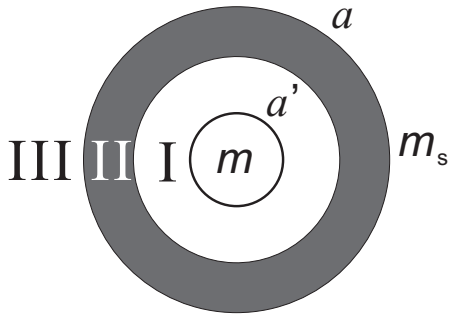}
\end{center} \caption{\label{fig3} The configuration of the dynamical
case for one shell in the comoving coordinates.   $m$ and  $m_s $
 are the total gravitating masses of the black hole
and the shell, respectively. $a'$
 and  $a$ are the radii of the inner and outer
boundaries of the shell in the comoving coordinates, respectively.}
\end{figure}

Fig. \ref{fig3} is the configuration of the one shell case.
To investigate the dynamical solution, we follow the method in
\cite{14}, i.e. obtaining the solution in the comoving coordinates
first. The general form of the metric is \be{eq15} ds^2  = d\tau ^2
- e^{\bar \omega } dR^2  - e^\omega  (d\theta ^2  + \sin ^2 \theta
d\phi ^2 ), \ee where  $\bar \omega (R,\tau )$
 and  $\omega (R,\tau )$
 are the unknown functions. In the
comoving coordinates, the only non-zero component of the energy
momentum tensor is  $T_4^4  = \rho $. As the results in \cite{14},
we can obtain the following equations from the field equation
\be{eq16}e^\omega   = (F\tau  + G)^{4/3},\ee
 \be{eq17}e^{\bar \omega }  =
e^\omega  \omega '^2 /4, \ee
 \be{eq18}8\pi \rho  = \frac{4}{3}\;(\tau
+ G/F)^{ - 1} (\tau  + G'/F')^{ - 1},\ee where $F$
  and  $G$ are
arbitrary functions of $R$. Dot and prime are the partial
differentiation with respect to $\tau $
  and  $R$, respectively. At  $\tau=0$, we
obtain \be{eq19}F =  - \sqrt {6\pi \rho _0 (G^2  + C_1 )}. \ee

As in \cite{14}, we choose  $G = R^{3/2}$.

In region I and III,  $\rho _0  = 0$, therefore,  $F =
{\rm{constant}}$. We will determine this constant and $C_1$ by the
matching condition in the next paragraph. As obtained in \cite{14},
in region III $F_{\rm III}= - \frac{3}{2}\sqrt {r_0 } $, where $r_0
= 2M$. According to the well-known Birkhoff's theorem for
spherically symmetric mass distribution, $M$ here is the total
gravitating mass of the system, i.e., including rest mass and
gravitational energy.

Throughout this paper, our matching conditions are chosen such that
the metric is continuous across the boundaries between the different
regions. In Sec. \ref{sec3}, we will verify that in our solutions
found this way the extrinsic curvature is also continuous at the
boundaries. Since $F_{\rm III}= - \frac{3}{2}\sqrt {r_0 }$, to
assure the metric is continuous at the boundary between region III
and II, i.e. $F_{\rm III}(R=a)=F_{\rm II}(R=a)$, according to Eq.
(\ref{eq19}), we obtain the value of $C_1 $ as $C_{1}=\frac{3}{{4\pi
\rho _0 }}M - a^3 $. Therefore, to assure the metric is continuous
at the boundary between region II and I, i.e. $F_{\rm
II}(R=a')=F_{\rm I}(R=a')$, the value of $F$ in region I should be $
 F_{\rm I}=- \frac{3}{2}\sqrt {2[M - \frac{4}{3}\pi \rho _0 (a^3  - a'^3 )]}\equiv
 - \frac{3}{2}\sqrt {2m}\equiv - \frac{3}{2}\sqrt {r_0'}
$ (we also denote $m_s\equiv \frac{4}{3}\pi \rho _0 (a^3 - a'^3 )$).
In Sec. \ref{sec212}, we will show that $m$ and $m_s$ are the total
gravitating masses of the central black hole and the shell,
respectively.

For the spherically symmetric ball case (just as the solution in \cite{14}), $C_1 $  must
be $0$. However, in the case that a shell collapses towards a pre-existing black hole,
$C_1 $ can be any value. For example, $C_1<0 $ means that the mass of the central black hole
is smaller than the mass of the inner region interior to the shell in the spherically symmetric
ball case. Thus, the gravitational force on the inner layers of the shell will become
so weak that the inner layer will take more time than the outer layers (it can be seen
from Eq. [\ref{eq21}]) to arrive at $r = 0$ \cite{17}. Actually, it means the crossing
of the layers will happen. We will not discuss the crossing case in this paper, since
no comoving coordinates exist in that case. This means that we will only study the cases when $C_1\ge0$.
Therefore, our result cannot be compared
directly with the result obtained for a collapsing shell (without a black hole in the
center, i.e., $C_1<0$) either without any thickness \cite{20} or only with first-order thickness \cite{21}.

Up to now we have obtained the solution in the comoving coordinates.
Then we try to transform the solution into the ordinary coordinates,
in which the metric has the form as \be{eq20} ds^2  = B(r,t)dt^2  -
A(r,t)dr^2  - r^2 (d\theta ^2  + \sin ^2 \theta d\phi ^2 ). \ee

The transformation of $r$ is obvious. We must choose
\be{eq21}
r=e^{\omega /2}  = (F\tau  + G)^{2/3}.
\ee

Using the contravariant form of the metric and requiring that the $g^{tr}$ term
vanishes, we have

\be{eq22}
t'/\dot t = \dot rr'.
\ee

Substituting the expressions of $r$ in three regions into Eq.
(\ref{eq22}), we can obtain three partial differential equations of
$t(R,\tau )$.

In region III, the solution of Eq. (\ref{eq22}) is
\be{eq23} t =
L(x),\;x = \frac{2}{{3\sqrt {r_0 } }}(R^{3/2}  - r^{3/2} ) - 2\sqrt
{rr_0 } + r_0 \ln \frac{{\sqrt r  + \sqrt {r_0 } }}{{\sqrt r  -
\sqrt {r_0 } }}. \ee

In region II, we note the solution of \be{eq24} t'/\dot t =  -
\frac{2}{3}\sqrt R \;[R^{3/2}  - \tau \sqrt {6\pi \rho _0 (R^3  +
C_1 )} ]^{ - \frac{2}{3}}[\sqrt {6\pi \rho _0 (R^3 + C_1 )} - 6\pi
\rho _0 R^{3/2} \tau ] \ee as $t = M(y)$, where $y$ is a function of
$R$ and $\tau$.

The characteristic  equation of Eq. (\ref{eq24}) is

\be{eq25} \frac{{d\tau }}{{dR}} = \frac{2}{3}\sqrt R \;[R^{3/2}  -
\tau \sqrt {6\pi \rho _0 (R^3  + C_1 )} ]^{ - \frac{2}{3}}[\sqrt
{6\pi \rho _0 (R^3  + C_1 )}  - 6\pi \rho _0 R^{3/2} \tau ] . \ee

In region I, the solution of Eq. (\ref{eq22}) is \be{}t = N(x'),\;x'
= \frac{2}{{3\sqrt {r_0 '} }}(R^{3/2}  - r^{3/2} ) - 2\sqrt {rr_0 '}
+ r_0 '\ln \frac{{\sqrt r  + \sqrt {r_0 '} }}{{\sqrt r  - \sqrt {r_0
'} }}.
 \ee
where $L$, $M$, and $N$  are arbitrary functions of their
arguments. Since the metric in the ordinary coordinates in region
III is the Schwarzschild metric, $L(x)$ should be $x$.

\subsubsection{\bf{Analytic solution when $C_1=0$}}\label{sec211}

  Eq. (\ref{eq24}) can be solved analytically only when $C_1=0$ and
the results are \be{}t = M(y), y = \frac{1}{2}(R^2 - a^2 ) +
\frac{1}{k}(1 - \frac{3}{2}\sqrt k \tau )^{\frac{2}{3}},
 \ee
where $k = \frac{{8\pi \rho _0 }}{3}$.

Then if we require $L=M$ at $R = a$
  for any  $\tau $, the form of $M$ must be
\be{eq28} M(y) = \frac{2}{{3\sqrt {r_0 } }}[a^{\frac{3}{2}}  -
(kay)^{3/2} ] - 2\sqrt {kayr_0 }  + r_0 \ln \frac{{\sqrt {kay}  +
\sqrt {r_0 } }}{{\sqrt {kay}  - \sqrt {r_0 } }}, \ee where  $y =
f(x')$ is determined by the following equation,

\begin{eqnarray}
x' =\frac{2}{{3\sqrt {r_0 '} }}[a'^{\frac{3}{2}} - (ka')^{3/2} (y -
\frac{1}{2}a'^2  + \frac{1}{2}a^2 )^{3/2} ] - 2\sqrt {r_0 'ka'\;[y -
\frac{1}{2}(a'^2  - a^2 )]}  \nonumber\\ +  r_0 '\ln \frac{{\sqrt
{ka'\;[y - \frac{1}{2}(a'^2  - a^2 )]}  + \sqrt {r_0 '} }}{{\sqrt
{ka'\;[y - \frac{1}{2}(a'^2  - a^2 )]}  - \sqrt {r_0 '} }}.
\end{eqnarray}

Using the relations \be{}g^{rr}  =  - (1 - \dot r^2 ), \ee and
\be{}{g}^{tt}  = \dot t^2 (1 - \dot r^2 ), \ee we can obtain the
metric in the ordinary coordinates

In region III, \be{}g_{tt}  = 1 - r_0 /r, \ee \be{} g_{rr} = - (1
- r_0 /r)^{ - 1}.\ee

In region II,

\be{eq34}
g_{rr}  =  - (1 - \frac{{kR^3 }}{r})^{ - 1}.
\ee

Since \be{eq35} \dot t = \frac{{a^{5/2} k^2 y^{\frac{3}{2}}
}}{{\sqrt {r_0 } (aky - r_0 )(1 - \frac{3}{2}\sqrt k \tau
)^{\frac{1}{3}} }}, \ee

therefore,
\be{eq36}
 g_{tt}  = \frac{{r_0 (aky - r_0 )^2 (1 -
\frac{3}{2}\sqrt k \tau )^{\frac{2}{3}} }}{{a^5 k^4 y^3 }}(1 -
\frac{{kR^3 }}{r})^{-1}, \ee where
 $k \equiv \frac{8}{3}\pi
\rho_0$.

In region I, in the ordinary coordinates, we can obtain \be{eq37}
g_{tt} = h(t)(1 + C_4 /r), \ee \be{eq38} g_{rr }  =  - (1 + C_4
/r)^{ - 1} .\ee

If we require $g_{rr} $
  and $g_{tt}$ are continuous at the inner boundary of the shell with
Eq. (\ref{eq34}) and (\ref{eq36}), respectively, we have
\be{eq39}C_4 = - r_0 '
 \ee
 and
 \be{eq40}
h(t) =\frac{{(1 - \frac{3}{2}\sqrt k \tau )^2 }}{{[(1 -
\frac{3}{2}\sqrt k \tau )^{\frac{2}{3}}  - r_0 '/a']^2
}}\frac{{\{[\frac{1}{2}(a'^2  - a^2 ) + \frac{1}{k}(1 -
\frac{3}{2}\sqrt k \tau )^{\frac{2}{3}} ] - a^2 \}^2
}}{{k[\frac{1}{2}(a'^2  - a^2 ) + \frac{1}{k}(1 - \frac{3}{2}\sqrt k
\tau )^{\frac{2}{3}} ]^3 }},
 \ee
where $\tau  = \tau (t)$
  is determined  by the following equation
\be{} t = \frac{2}{{3\sqrt {r_0 } }}[a^{\frac{3}{2}} - (kay(\tau
))^{3/2} ] - 2\sqrt {kay(\tau )r_0 }  + r_0 \ln \frac{{\sqrt
{kay(\tau )}  + \sqrt {r_0 } }}{{\sqrt {kay(\tau )} - \sqrt {r_0 }
}} . \ee

\subsubsection{\bf{Numerical solution when $C_1>0$}}\label{sec212}

\begin{figure*}
\begin{center}
\includegraphics[width=0.47\textwidth]{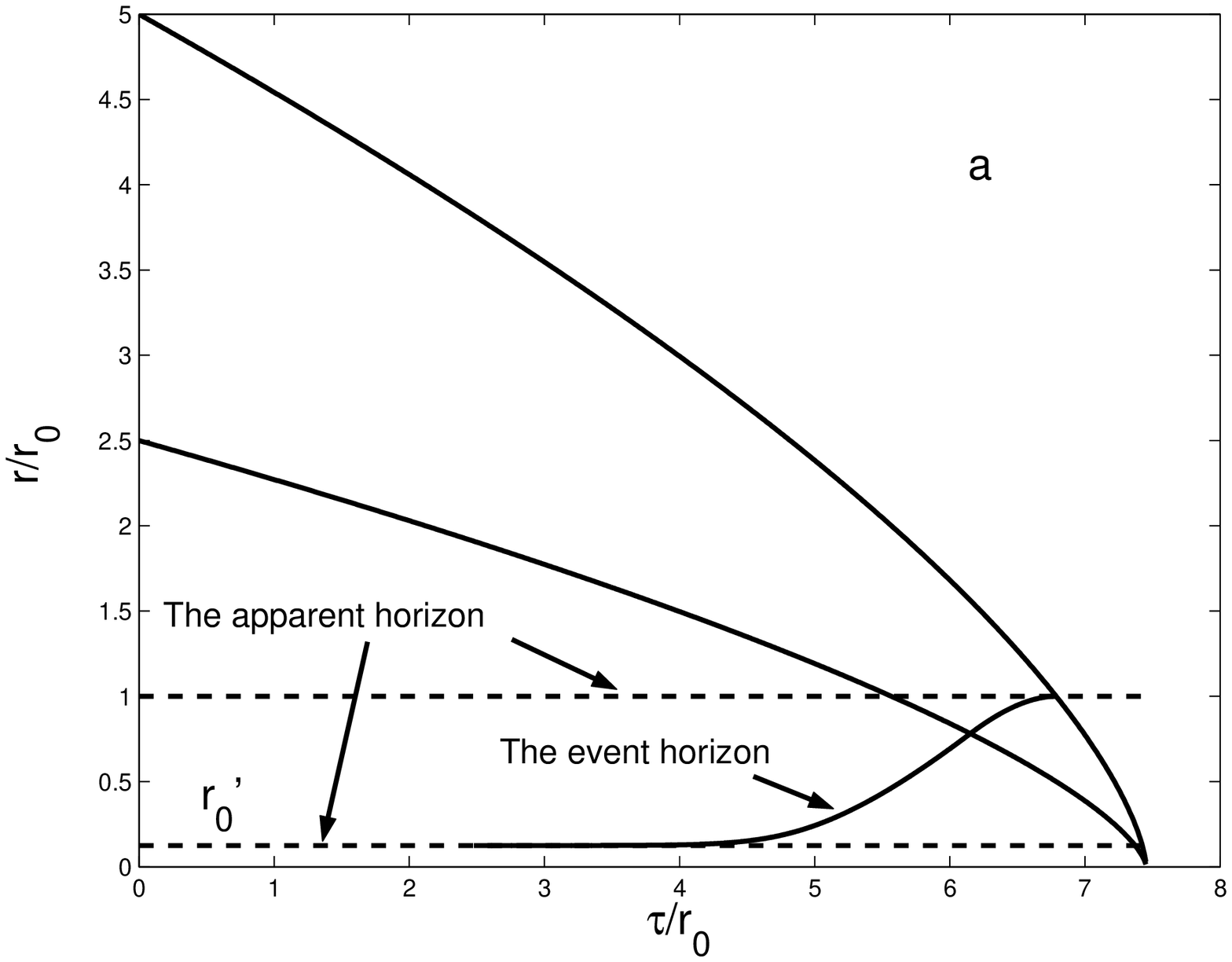}
\hspace{0.5cm}
\includegraphics[width=0.47\textwidth]{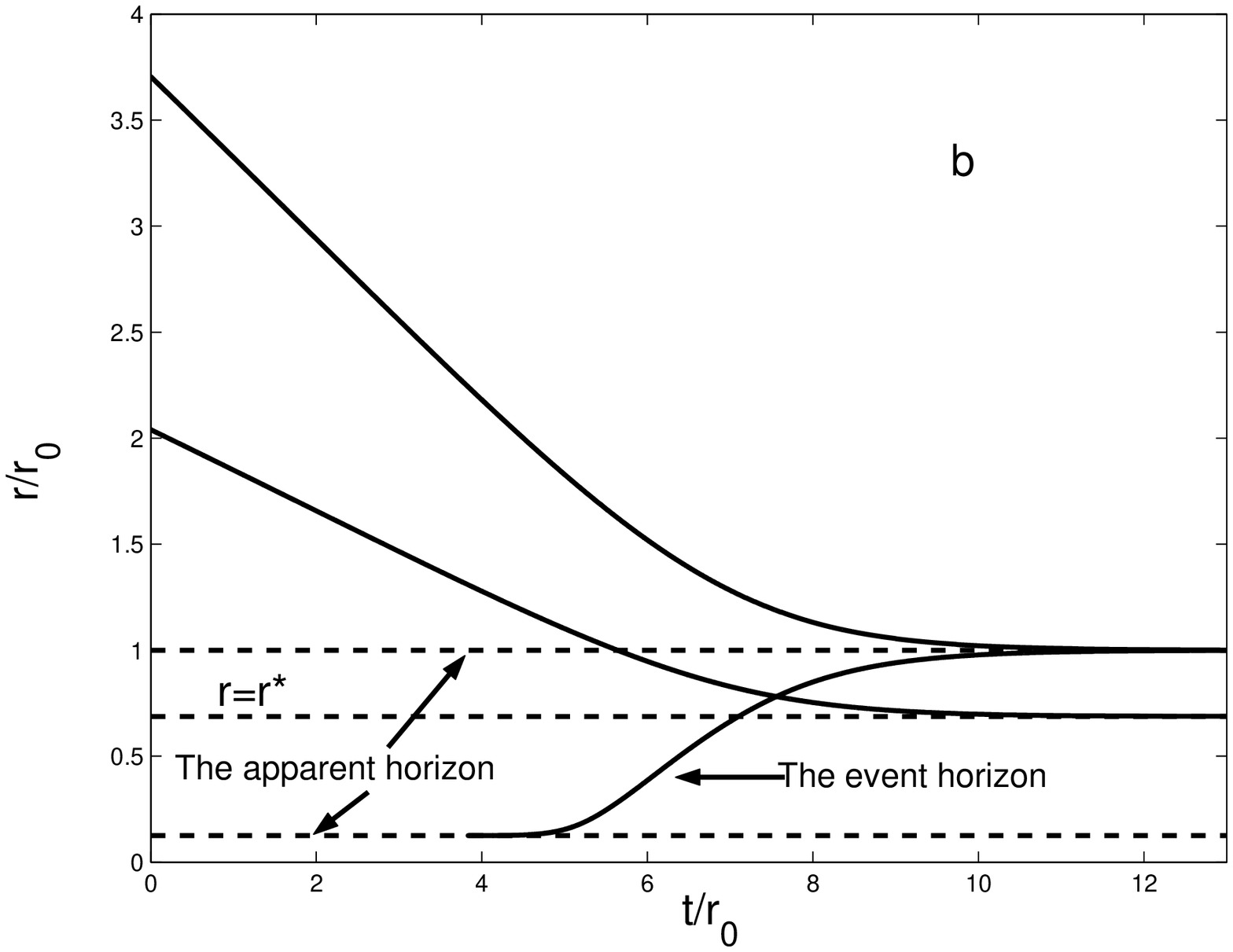}
\includegraphics[width=0.47\textwidth]{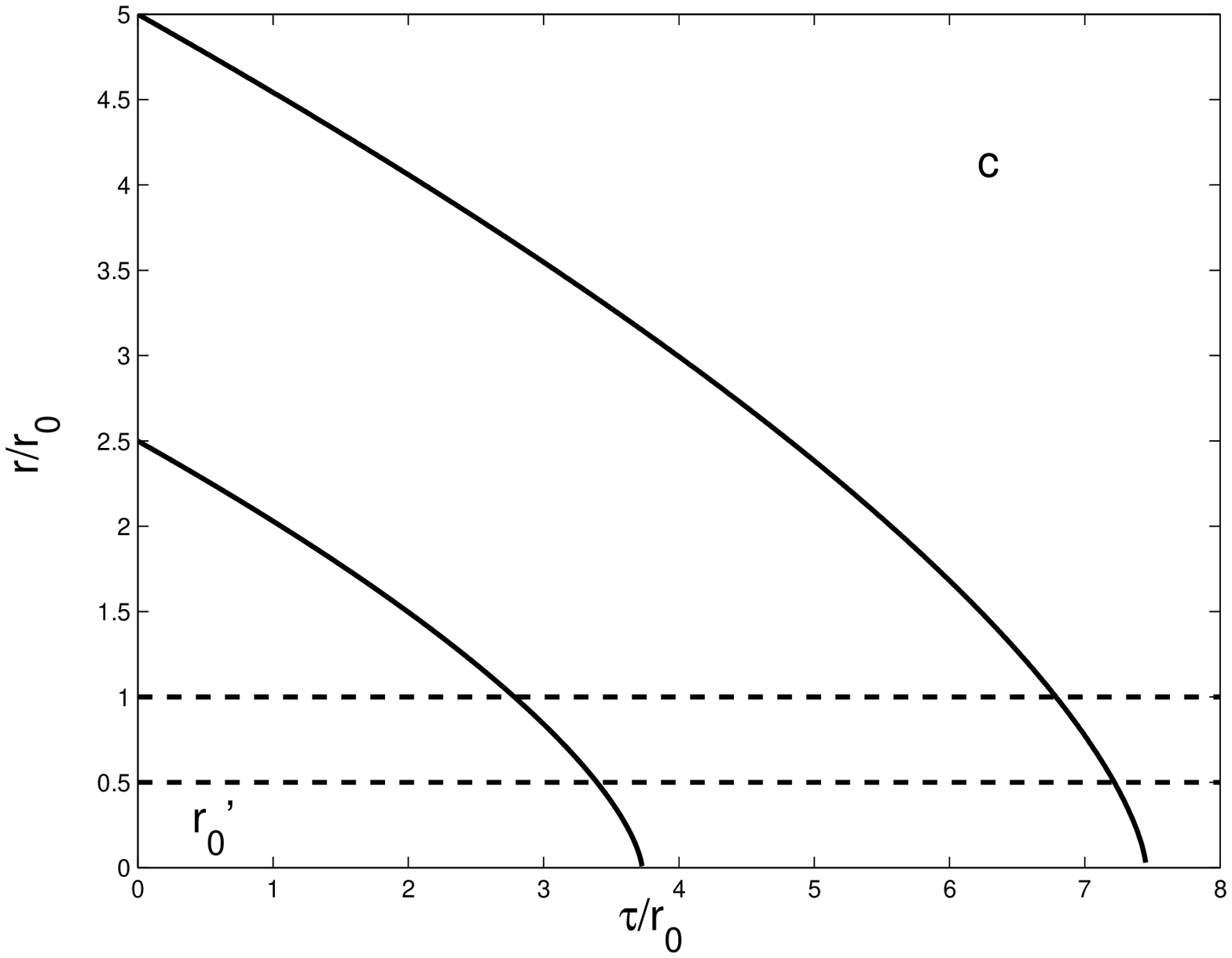}
\hspace{0.5cm}
\includegraphics[width=0.47\textwidth]{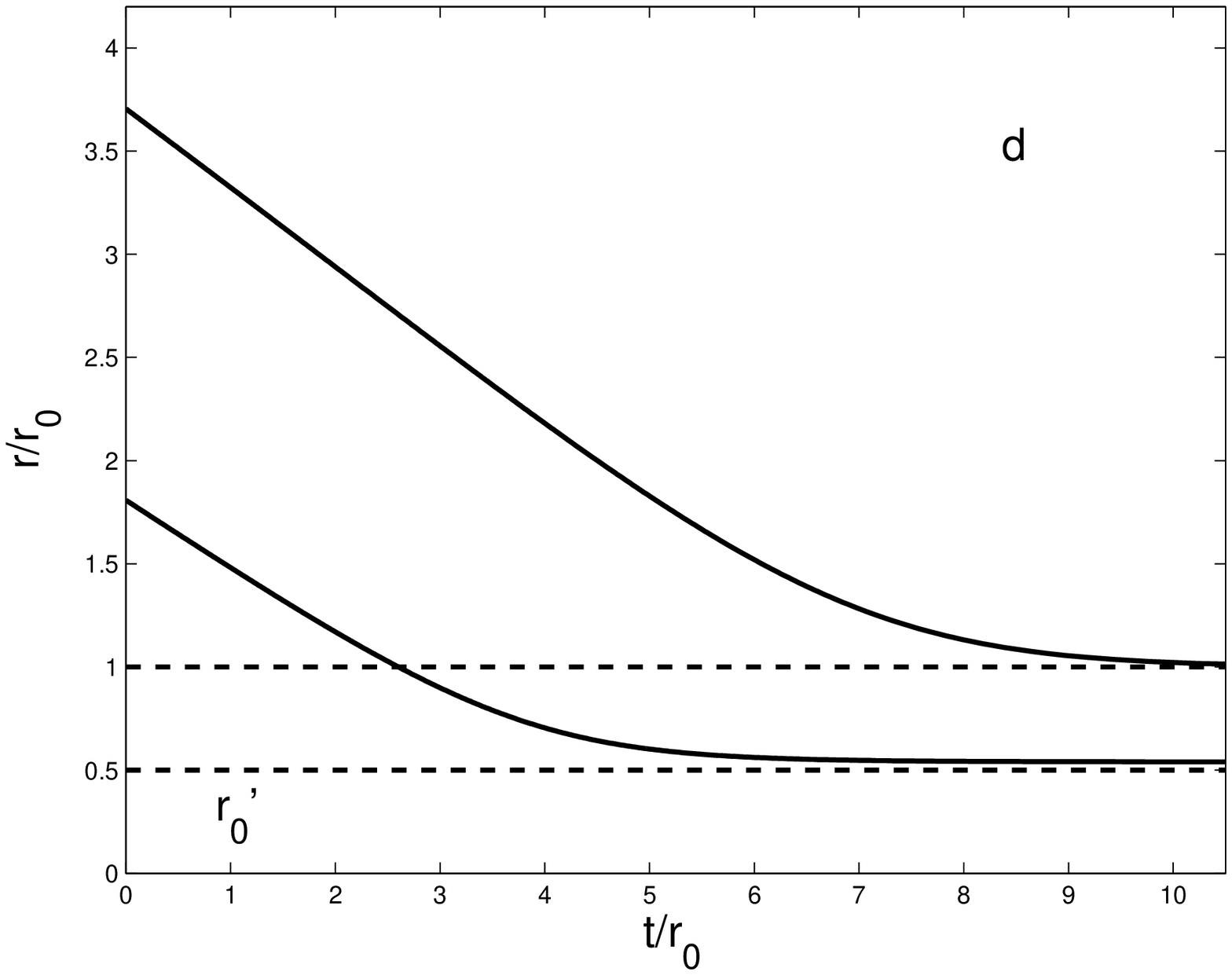}
\end{center}
 \caption{\label{fig4} The dynamical solutions for one
shell case. (a) and (b) are evolution curves for  $a = 5r_0 $,  $a'
= 2.5r_0 $, and $r_0 ' = 1/8r_0 $ (i.e. $C_1=0$) with comoving time
and coordinate time, respectively. The evolution of the event and
apparent horizons are also shown. (c) and (d) for  $a = 5r_0 $, $a'
= 2.5r_0 $, and $r_0 ' = 0.5r_0 $ (i.e. $C_1>0$) with comoving time
and coordinate time, respectively.}
\end{figure*}

In general cases, Eq. (\ref{eq24}) can be solved numerically only.
From Eq. (\ref{eq23}), the value of $t$ at the outer boundary of the
shell can be determined. It can be easily shown that the value of
$t$ along every curve determined by Eq. (\ref{eq25}), i.e.  $\tau  =
\tau (R)$, is constant. Therefore, using the value of $t$ from Eq.
(\ref{eq23}) as the boundary value, we integrate Eq. (\ref{eq25})
along the characteristic lines and obtain how the shell evolves with
coordinate time $t$. In Fig. \ref{fig4}, we show the evolution
curves for the cases  $C_1  = 0$ and  $C_1  > 0$. As it can be seen,
the inner boundary of the shell crosses the Schwarzschild radius
with finite time and approaches to a position  $r* = \frac{3}{2}r_0
^{\frac{2}{3}} r_0 '^{\frac{1}{3}} - \frac{1}{2}r_0 '$ (This is
obtained from the transformations Eq. [\ref{eq21}] and Eq.
[\ref{eq28}]), whereas the outer boundary approaches to the
Schwarzschild radius. (However, if there is another shell outside
this shell, the outer boundary of this shell will also cross the
Schwarzschild radius with finite time, as we will show in Sec.
\ref{sec22})

As shown in Fig. \ref{fig4} the event horizon of the system
increases just from the location $r=2m\equiv r_0'$. Therefore,
$r=r_0'$ corresponds to the event horizon of the central black hole
and $m$ is the total gravitating mass of the central black hole. As a result,
$m_s$ can be recognized as the total gravitating mass of the shell.

\begin{figure}
\begin{center}
\includegraphics[width=0.5\textwidth]{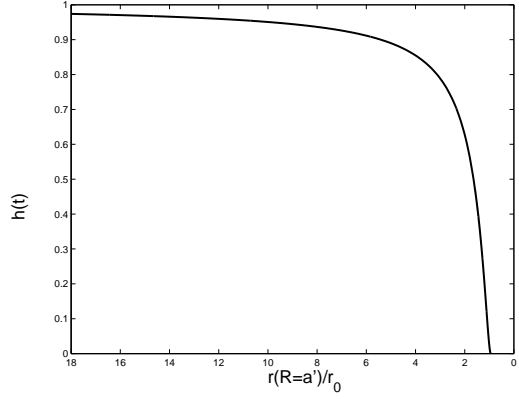}
\end{center}
\caption{\label{fig5} The evolution of $h(t)$
  with the position of the outer
boundary of the shell for the case  $a = 5r_0 $,  $a' = 2.5r_0 $,
and $r_0 ' = 1/8r_0 $.}
\end{figure}

We find that the metric in region I is not the standard
Schwarzschild metric; actually, it should be a function of time in
the dynamical case. The evolution curve of  $h(t)$ is shown in Fig.
\ref{fig5}. When the shell is far from the central black hole, the
value of $h(t)$  is almost equal to $1$, i.e. the metric is
approximately the standard Schwarzschild solution. When the shell is
near the horizon, the value of $h(t)$
  decreases rapidly and
approaches to $0$, i.e. the clock in region I becomes slower and
slower. It does not contradict the Birkhoff's theorem, since in the
proof of the Birkhoff's theorem the region outside the central mass
is vacuum \cite{4}. Therefore, in this case we could adjust the
clock to ensure that metric asymptotically approaches to the
Minkowski metric at infinity. However, in our problem there is a
shell, but not the vacuum outside the region I. Thus we should
ensure that the metric is continuous at the boundary between region
I and the shell and therefore cannot arbitrarily select the clock
inside the shell, i.e. region I.

\subsection{The solution for two shells}\label{sec22}
\begin{figure}
\begin{center}
\includegraphics[width=0.4\textwidth]{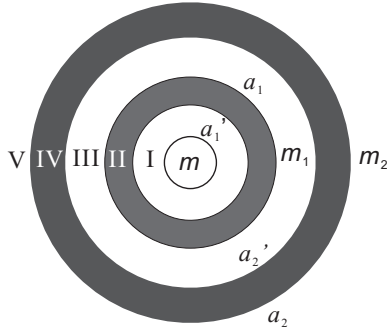}
\end{center} \caption{\label{fig6} The configuration of the dynamical
case for two shells in comoving coordinates.  $m$,   $m_1 $ and $m_2
$ are the total gravitating masses of the black hole, the shell 1
and 2, respectively. $a_1 '$ and  $a_1 $
 are the radii of the
inner and outer boundaries of shell 1, respectively.  $a_2 '$
 and  $a_2$
 are the radii of the
inner and outer boundaries of shell 2, respectively.}
\end{figure}

In this section, we investigate the solution for two shells. The
configuration of this problem is shown in Fig. \ref{fig6}. The
process to obtain the solution is similar to the one shell case,
however, there are five regions. The result of $F$ is
 \be{} F = \left\{ {\begin{array}{*{20}c}
   { - \frac{3}{2}\sqrt {r_0 '} \;\;\;\;\;\;\;\;\;\;;R < a_1 '}  \\
   { - \frac{3}{2}\sqrt {\frac{{8\pi \rho _1 }}{3}(R^3  - a_1 ^3 ) + r_0 ^{\prime \prime } } \;\;\;;a_1 ' < R < a_1 }  \\
   { - \frac{3}{2}\sqrt {r_0 ^{\prime \prime } } \;\;\;;a_1  < R < a_2 '}  \\
   {\begin{array}{*{20}c}
   { - \frac{3}{2}\sqrt {\frac{{8\pi \rho _2 }}{3}(R^3  - a_2 ^3 ) + r_0 } \;\;\;;a_2 ' < R < a_2 }  \\
   { - \frac{3}{2}\sqrt {r_0 } \;\;\;;R > a_2}  \\
\end{array}}  \\
\end{array}} \right.,
\ee where  $r_0  = 2M,\;r_0 ' = 2m$, $r_0 ^{\prime \prime }  = 2(m
+ m_1 )$.

The form of the transformation in region V is the same as that in
region III of the one shell case. The transformations for other
regions are shown below.

In region IV, \be{eq43} t'/\dot t =  - \frac{2}{3}\sqrt R \;[R^{3/2}
- \tau \sqrt {6\pi \rho _2 (R^3  + C_2 )} ]^{ - \frac{2}{3}} [\sqrt
{6\pi \rho _2 (R^3  + C_2 )}  - 6\pi \rho _2 R^{3/2} \tau ]. \ee

The characteristic equation of Eq. (\ref{eq43}) is \be{}
\frac{{d\tau }}{{dR}} = \frac{2}{3}\sqrt R \;[R^{3/2}  - \tau \sqrt
{6\pi \rho _2 (R^3  + C_2 )} ]^{ - \frac{2}{3}} [\sqrt {6\pi \rho _2
(R^3  + C_2 )}  - 6\pi \rho _2 R^{3/2} \tau ] , \ee where  $C_2 =
\frac{3}{{4\pi \rho _2 }}M - a_2 ^3 $.

In region III, \be{} t = O(x''),\;x'' = \frac{2}{{3\sqrt {r_0 ''}
}}(R^{3/2}  - r^{3/2} ) - 2\sqrt {rr_0 ''} + r_0 ''\ln \frac{{\sqrt
r  + \sqrt {r_0 ''} }}{{\sqrt r - \sqrt {r_0 ''} }} ,\ee where
$O(x'')$
 is an arbitrary function
of  $x''$.

In region II, \be{eq46} t'/\dot t =  - \frac{2}{3}\sqrt R \;[R^{3/2}
- \tau \sqrt {6\pi \rho _1 (R^3  + C_1 )} ]^{ - \frac{2}{3}}[\sqrt
{6\pi \rho _1 (R^3  + C_1 )}  - 6\pi \rho _1 R^{3/2} \tau ] . \ee

The characteristic equation of Eq. (\ref{eq46}) is \be{}
\frac{{d\tau }}{{dR}} = \frac{2}{3}\sqrt R \;[R^{3/2} - \tau \sqrt
{6\pi \rho _1 (R^3 + C_1 )} ]^{ - \frac{2}{3}}[\sqrt {6\pi \rho _1
(R^3  + C_1 )}  - 6\pi \rho _1 R^{3/2} \tau ] ,\ee where  $C_1  =
\frac{3}{{4\pi \rho _1 }}(m + m_1 ) - a_1 ^3 $.

In region I, \be{} t = P(x'),\;x' = \frac{2}{{3\sqrt {r_0 '}
}}(R^{3/2}  - r^{3/2} ) - 2\sqrt {rr_0 '}  + r_0 '\ln \frac{{\sqrt r
+ \sqrt {r_0 '} }}{{\sqrt r  - \sqrt {r_0 '} }},\ee where $P(x')$ is
an arbitrary function of $x'$.

\begin{figure*}
\begin{center}
\includegraphics[width=0.47\textwidth]{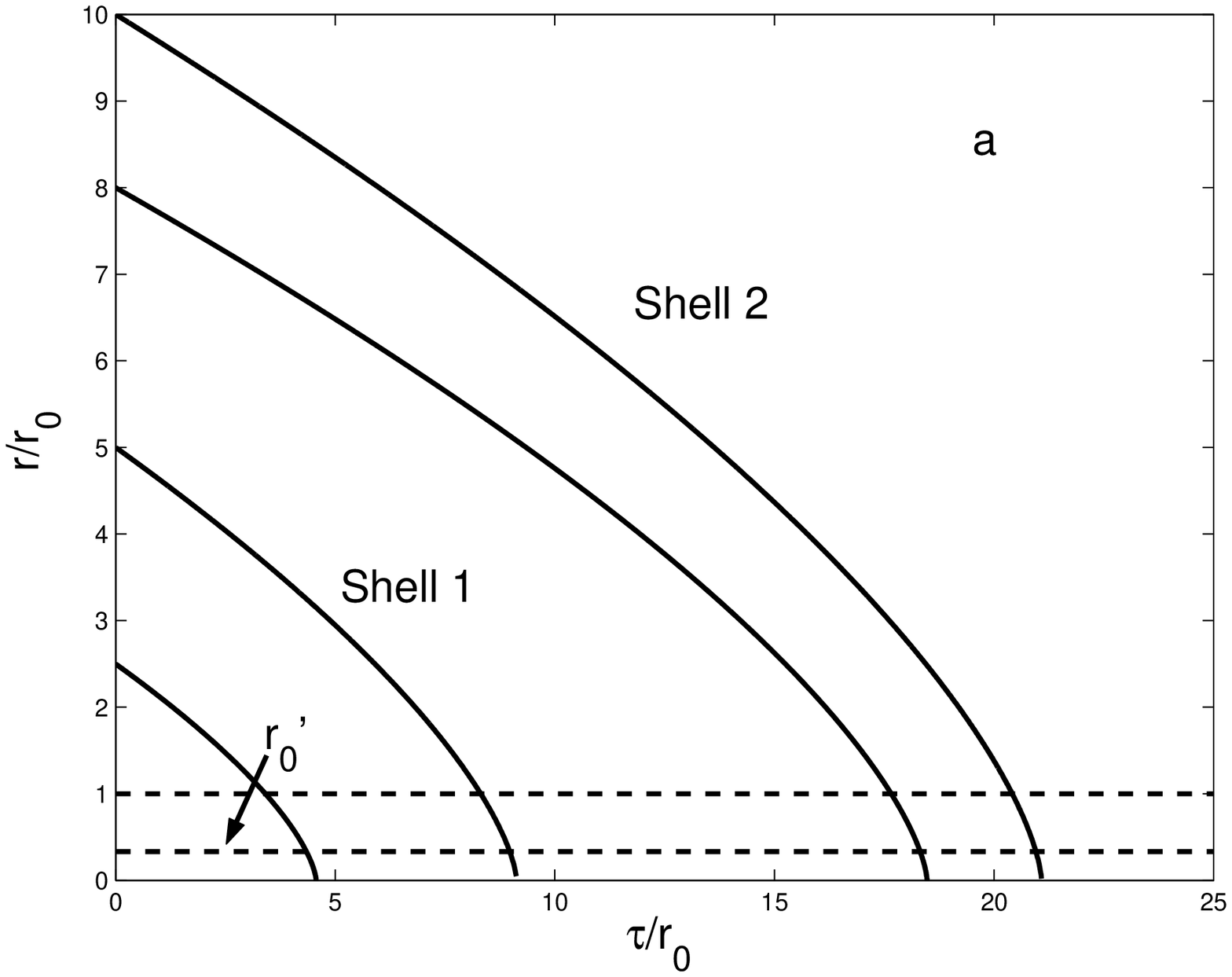}
\hspace{0.5cm}
\includegraphics[width=0.47\textwidth]{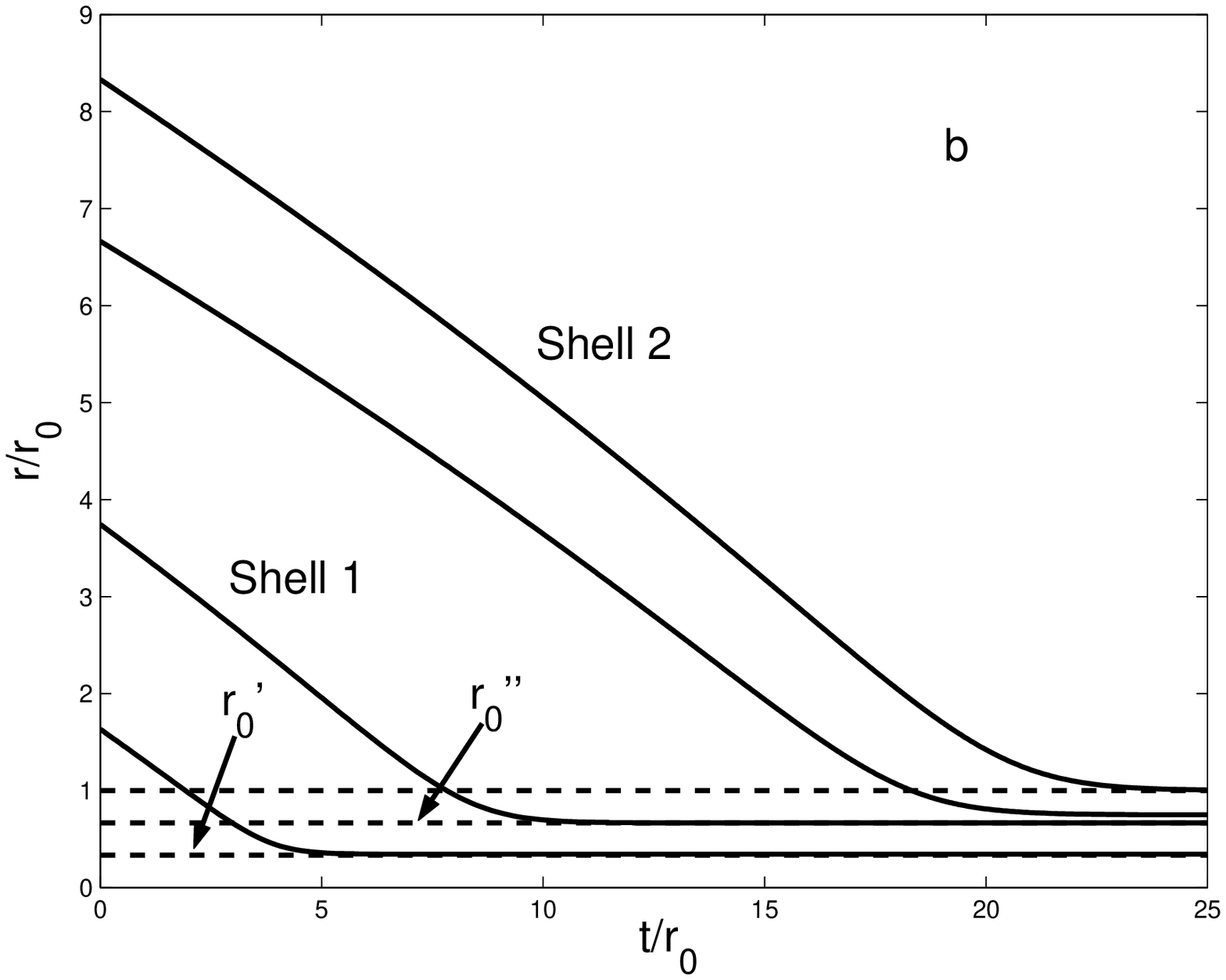}
\end{center} \caption{\label{fig7} The dynamical solutions for two
shells case. (a) and (b) are evolution curves for  $a_2  = 10r_0 $,
$a_2 ' = 8r_0 $,  $a_1 = 5r_0 $, $a_1 ' = 2.5r_0 $, $r_0 ' = 1/3r_0
$, and $r_0 '' = 2/3r_0 $
 with
comoving time and coordinate time, respectively.}
\end{figure*}

\begin{figure}
\begin{center}
\includegraphics[width=0.5\textwidth]{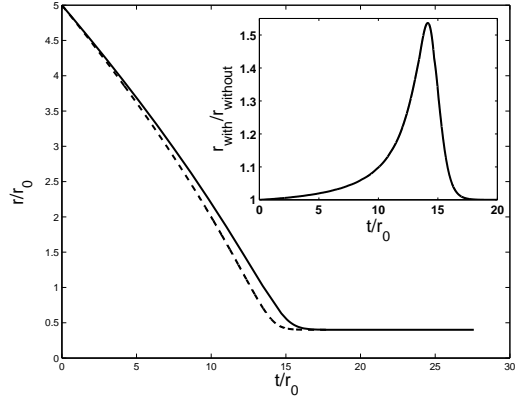}
\end{center} \caption{\label{fig8}  The comparison of the evolution
of the outer boundary of shell 1 between the case with (solid) or
without (dashed) shell 2. The parameters of the shells are  $a_2 =
10r_0 $, $a_2 ' = 6r_0 $, $a_1 ' = 2.5r_0 $,  $a_1  = 5r_0 $, $r_0 '
= 1/5r_0 $, and  $r_0 '' = 2/5r_0 $, where $r_0$ is the
Schwarzschild radius corresponding to the total gravitating mass of
the system, i.e., the sum of the gravitating masses of the
pre-existing black hole, shell 1 and shell 2. The inset is the ratio
of the solid line to the dashed line. It is seen clearly that
outside matter, though spherically symmetric, does influence the
motion of matter inside.}
\end{figure}

To obtain the evolution curves of the shells, we solve the above
equations numerically with the same method in Sec. \ref{sec212}.
Fig. \ref{fig7} shows the evolution curves of two shells. The outer
boundary of shell 1 crosses the Schwarzschild radius corresponding
to the total gravitating mass within finite coordinate time. In Fig.
\ref{fig8}, we show the evolution curves of the outer boundary of
shell 1 in the case with or without shell 2. As we have pointed out
in Sec. \ref{sec21}, the clock inside the shell is slower, therefore
the outer boundary of shell 1 will take more time to approach the
asymptotic line in the case with shell 2.

\section{Discussions and Conclusions}\label{sec3}

In this work, our matching conditions selection requires that the
metric is continuous at the boundaries. Here we verify that in our
solutions found this way the extrinsic curvature is also continuous
at each boundary (see the detailed discussion about the extrinsic
curvature in \cite{23}). Since the boundary is defined as
$R=constant$, the normalized normal vector is

\be {} n_\alpha   = (0,\frac{1}{{\left| {g^{RR} } \right|^{1/2} }},0,0). \ee The definition of extrinsic
curvature is $K_{ab}  = n_{\alpha ;\beta } e_a^\alpha  e_b^\beta  $. We choose $y^a  = (\tau ,\theta ,\varphi )$
as the coordinates on the boundary. Therefore, the non-vanishing components of $K_{ab} $ are

\be{}
 K_{\tau \tau }  = n_{\tau ;\tau }  =  - \Gamma _{\tau \tau }^R
n_R = \frac{1}{2}n_R g^{RR} g_{\tau \tau ,R}  = 0, \ee

\be{} K_{\theta \theta }  = n_{\theta ;\theta }  =  - \Gamma
_{\theta \theta }^R n_R  = \frac{1}{2}g^{RR} g_{\theta \theta ,R}
n_R  = \left( {F\tau  + G} \right)^{2/3}, \ee

\be{}
 K_{\varphi \varphi }  = n_{\varphi ;\varphi }  =  - \Gamma
_{\varphi \varphi }^R n_R  = \frac{1}{2}g^{RR} g_{\varphi \varphi
,R} n_R  = \left( {F\tau  + G} \right)^{2/3} \sin ^2 \theta. \ee{}
Obviously, they are all continuous at all boundaries.

With the dynamical solutions presented in this work, we conclude
that

 (a) In general relativity, the outer mass
influences the inner space-time even if the mass distribution is spherically symmetric;

(b) The metric inside the shell can not be the standard Schwarzschild form and is time
dependent in the dynamical case;

(c) A dust shell of finite thickness can indeed cross the black hole's event horizon, as observed by an external
observer, except for its outer surface. This means that asymptotically only an infinitesimal amount of matter
remains just outside the event horizon. Therefore such a ``star" is qualitatively different from the so-called
the ``frozen star", which supposedly holds all in-falling matter outside its event horizon;

(d) In the two-shell case, even the outer boundary of the inner shell can be observed by an external observer to
collapse into the Schwarzschild radius, thus no matter can be accumulated outside a black hole's event horizon.
Since we can mimic the outer shell as all matter between the observer and the in-falling shell being observed,
we can conclude that the outside observer indeed ``sees" the inner shell falls into the black hole completely.
We note that the outer layer of matter does not necessarily block the view of the outside observer to the inner
layer, since the outer layer can be optically thin to at least a certain wavelength of electromagnetic
radiation, or is made of dark matter and is thus completely transparent to electromagnetic radiation. Even if
the outer layer is optically thick completely, such as the case when the fall-back matter approaches to the
black hole after the core collapse of a massive star, neutrinos can still find their way out and photons may
eventually escape after a lengthy radiative transfer process.

In our solution, the metric between the shell and the black hole
should not be the standard Schwarzschild form, and there should be
an additional time-dependent factor $h(t)$ in $g_{tt} $ (please
refer to equations Eq. (\ref{eq37}) and  (\ref{eq40}), Fig.
\ref{fig5}. This factor $h(t)$ decreases from 1 to 0 when the shell
collapses towards the Schwarzschild radius. This means that the
clock in the inner region of the shell is slower compared with the
case without the shell, and becoming even slower when the shell is
falling in. This effect may be testable experimentally in principle,
e.g., if a beam of light travels across the shell and a parallel
beam of light passes outside the shell (far enough from the shell),
then the two beams of light will travel at different velocities with
respect to the observers outside the shell. This previously
unrecognized effect is caused by the outer mass's influence to the
inner space-time; such effect does not exist in the Newtonian
Gravity. We could recognize it as the generalized Shapiro effect,
because this effect is due to the mass outside, but not inside as
the case of the standard Shapiro effect. We comment in passing that
in \cite{22}, the author directly identified the metric between two
thin shells as Schwarzschild metric and therefore his result can
recover to our one shell solution only when the mass of the outer
shell vanishes.

In the one shell case, the shell could almost cross the Schwarzschild radius except the outer boundary; such a
star is already qualitatively different from a ``frozen star". Nevertheless, a shell can indeed cross the
Schwarzschild radius in the two shell case, and the external observer can indeed observe a complete shell
falling into a black hole, since the outer shell in general cannot block the view to the inner shell completely,
as discussed above. In this sense we could observe the matter falls into a black hole and the ``frozen star"
paradox discussed in \cite{10} is naturally solved. As pointed out previously by us, the origin of the ``frozen
star" is the ``test particle" assumption for the in-falling matter, which neglects the influence to the metric
by the in-falling matter itself \cite{24}. In fact, for some practical astrophysical settings, the time taken
for matter falling into a black hole is quite short, even for the external observer \cite{24}. Therefore with
full and self-consistent general relativity calculation, we find that the in-falling matter will not accumulate
outside the event horizon, and thus the quantum radiation and Gamma Ray bursts predicted in \cite{11} and
\cite{12} are not likely to be generated. We predict that only gravitational wave radiation can be produced in
the final stage of the merging process of two coalescing black holes. Future simultaneous observations by X-ray
telescopes and gravitational wave telescopes shall be able to verify our prediction.

It is also interesting to note that, as can be seen from Fig.
\ref{fig4}b, \ref{fig4}d and \ref{fig7}b, in ordinary coordinates,
the matter will not collapse to the singularity  $(r = 0)$
 even with infinite coordinate time (if $r_0' = 0$, the inner boundary will take infinite time
  to arrive at  $r = 0$). It means that in real astrophysics sense, matter can never
arrive at the singularity (i.e., the exact center of the black
hole) with respect to the clock of the external observer.
Therefore, no gravitational singularity exists physically, even
within the framework of the classical general relativity.

As predicted by Hawking's area theorem, the event horizon grows
monotonously and continuously from $r_0 '$
  to $r_0 $
  when the shell collapses
towards the black hole, whereas the apparent horizon jumps to the
Schwarzschild radius when all the mass of the shell crosses into the
Schwarzschild radius (see Fig. \ref{fig4}a and 4b). Therefore, we
have presented a worked-out example of the area theorem.

\section*{Acknowledgments}
 The basic idea for this paper was sparked during several
stimulating discussions in May 2007 between SNZ and his former student Ms. Sumin Tang studying in Astronomy
Department, Harvard University (she also contributed partially to a progress report of this work published in a
conference proceedings \cite{24}). We also thank many discussions with Richard Lieu, Kinwah Wu, Kazuo Makishima,
Neil Gehrels, Masaruare Shibata, Ramesh Narayan, Zheng Zhao, Zonghong Zhu and Chongming Xu. SNZ also thanks Remo
Ruffini, Roy Kerr and Hernando Quevedo Cubillos for positive comments and encouragements on this work during the
6th Italian-Sino Workshop on Relativistic Astrophysics in ICRANet, Pescara, Italy. SNZ acknowledges partial
funding support by the Yangtze Endowment from the Ministry of Education at Tsinghua University, Directional
Research Project of the Chinese Academy of Sciences under project No. KJCX2-YW-T03 and by the National Natural
Science Foundation of China under project no. 10521001, 10733010, 10725313, and by 973 Program of China under
grant2009CB824800.

\end{document}